\documentclass[twocolumn,twocolappendix]{aastex63}
\usepackage{amsmath, graphicx, amssymb, amsfonts, hyperref}
\usepackage[utf8]{inputenc}
\usepackage{textcomp, bm}
\usepackage{lineno}

\received{December 11, 2019}
\revised{March 2, 2020}
\accepted{March 14, 2020}
\submitjournal{ApJ}

\shorttitle{Radiative Torque Disruption of Cometary Dust}
\shortauthors{Herranen, J.}

\graphicspath{{./}{figures/}}

\begin{document}
	
	\title{Rotational Disruption of Nonspherical Cometary Dust Particles by Radiative Torques}
	
	\correspondingauthor{Joonas Herranen}
	\email{joonas.herranen@iki.fi}
	
	\author[0000-0001-7732-9363]{Joonas Herranen}
	\affiliation{University of Helsinki}
	
	\begin{abstract}
		Rigorous statistical numerical analysis of the response of a nonspherical dust particle ensemble composed of aggregates of astronomical silicate is presented. It is found that the rotational disruption mechanism is not only likely to occur but to be a key element in explaining many separate observations of cometary dust. Namely, radiative torques are shown to spin-up and align cometary dust within the timescales of cometary activity. Additionally, the radiative torque alignment and disruption mechanisms within certain conditions are shown to be consistent with observations of rapid polarization of dust and spectral bluing of dust. The results indicate that radiative torques should be taken into account nearly universally when considering the evolution of cometary dust.
		
	\end{abstract}

	
	\section{Introduction} \label{sec:intro}
	Rotational disruption of bodies in different size regimes has been vigorously studied recently. For cometary objects, a particularly recent and interesting object was identified by \citet{Jewitt2019} to be tentatively the victim of torques due to outgassing instead of tidal disruption. Similar rotational disruption mechanism for small dust grains due to radiative torques \citep{Lazarian2007b}, or RATs, was recently proposed by \citet{Hoang2019}, provided large enough radiation intensities. Particularly, disruption in the case of submicron-scale dust occurs up to some parsecs away from type IA supernovae, which relate to radiation intensities at distances of some 1 au from the Sun. 
	
	\citet{Misconi1993} studied the rotational disruption of cosmic dust, particularly in the circumsolar case. In the study, rotation-inducing effects such as photonic and plasma interactions were found to cause disruption of dust particles inside a heliocentric radius of $ 10R_{\odot} $. Later on, in the review by \citet{Mann2004}, a reduction of particle population in the mass range of $ 10^{-14} $ to $ 10^{-12} $ g, roughly corresponding to particle size range 0.1 to 1 micron, occuring at heliocentric distances 0.1 to 1 au, was discussed. 
	
	Two observational missions, SOHO and STEREO, have detected many of the near-Sun comets known \citep{Battams2017, Knight2010, Jones2018} despite their main focus being solar environment analysis. During the perihelion passage, near-Sun comets are likely to reach temperatures high enough to sublimate the silicate dust material of the tail and the surface \citep{Mann2004}, or even cause, together with tidal effects, the catastrophic disruption of the whole comet \citep{Knight2014, Hui2015}. Sungrazing comets have been observed to exhibit changing polarization properties both at pre- and post-perihelion \citep{Thompson2015, Thompson2020}. 
	
	Finally, unusual increase in polarization degree and spectral bluing has been observed around the perihelion passage of comet 252P/LINEAR \citep{Kwon2019}. These measurements were attributed to the change in the dust population around the intensely heated nucleus of the comet. Namely, the evidence suggests that there is a lack of small fluffy particles in the dust population. 
	
	In the article, RATs and their role in the rotational disruption of cometary dust is studied for differently sized aggregate particles. The main scope of the article is to assess the magnitude of the effect of RATs, fundamentally based on exact numerical solutions of Maxwell's equations. While comets provide the main context, most of the assumptions on environments strive towards least loss of generality so that the results would provide useful ground rules for RAT spin-up and possible disruption. The implications of these results are reflected against the different observations discussed in the previous paragraphs. 
	
	\section{Radiative torque disruption mechanism}
	The RAT mechanism is based on the fact that helical dust particles have different responses to left- and right-handed polarizations of light. Angular momentum then transfers to helical scatterers, well described using analytic formulations of electromagnetic scattering by spheres \citep{Marston1984}, and successfully modeled for nonspherical particles by a geometric-optics-based analytical model \citep[AMO]{Lazarian2007b}, where the torques are due to specular reflection from a mirror, attached to a spheroidal parent body. 
	
	In planetary sciences, rotational motion induced by interactions between solar system bodies and electromagnetic radiation is often attributed to the YORP effect \citep{Rubincam2000}. Whereas YORP effect is related to anisotropic momentum exchange between the body and absorbed, reflected or emitted photons relative to the body center of mass, RATs emerge from direct exchange of angular momentum in the scattering interaction. Further, RATs are considered to be strictly a result of scattering, and they provide a component of the total torque, which includes e.g. the accelerating and decelerating contributions of emissions. In other words, when considering bodies, or rather particles, of sizes that are relative to the wavelength of light, a strict distinction between RAT and YORP effects is well-justified. 
	
	Given certain scattering conditions for a nonspherical particle, the resulting radiative torque may consist of components that spin up aligned particles \citep{Draine1996a}. Then, given enough time and low enough rotational dissipation, a critical angular velocity where the internal structure of the particle cannot hold against centrifugal forces may be reached \citep{Hoang2019}.
	
	Here, the above idea is applied to an ensemble of differently shaped and sized aggregate particles, for which the underlying scattering problem is solved using numerically exact volume-integral-equation-based $ T $-matrix solution \citep{Markkanen2016,Markkanen2012}.  The $ T $-matrix method is a generalization of the semianalytic Mie scattering theory. A $ T $-matrix is a linear map between vector spherical wave function expansions of the incident and scattered electromagnetic fields, first formulated by \citet{Waterman1965}. 
	
	Efficient and exact methods allow the analysis of the RAT response of a set of complex-shaped particles in a statistical manner. AMO-based analysed consider relevant RAT quantities as free parameters. In contrast, the ensemble analysis considers the same quantities as directly related to the accurate scattering solution realizing a statistical correspondence between a set of particle shapes and the scattering solution \citep{Herranen2019a}.
	
	\subsection{Dust model and critical angular speeds}

	The particle shapes are generated by ballistic cluster-cluster aggregating (BCCA) spheres. In BCCA, a new aggregate is produced by colliding two smaller sphere clusters generated in the previous step. After aggregation, each sphere is replaced with a different Gaussian random ellipsoid \citep{Muinonen2011}, defined by their axis ratio $ a:b:c = 1:0.8:0.9 $, standard deviation 0.125, and correlation length 0.35, which determine the statistics of the ellipsoid deformation.  A relatively low number of monomers are used to constrain the particle sizes, limited by contemporary computational limits. The resulting particles are thus a representative of a possible subset of realistic particles, ranging from small compact particles to slightly more fluffy larger aggregates. In situ observations at 67P/Churyumov-Gerasimenko \citep{Bentley2016} report a variety of compact single particles and large porous aggregates, shapes of which can at least approximately be sampled via the procedure above.
	
\begin{figure*}
		\centering
		\includegraphics[width=0.9\linewidth]{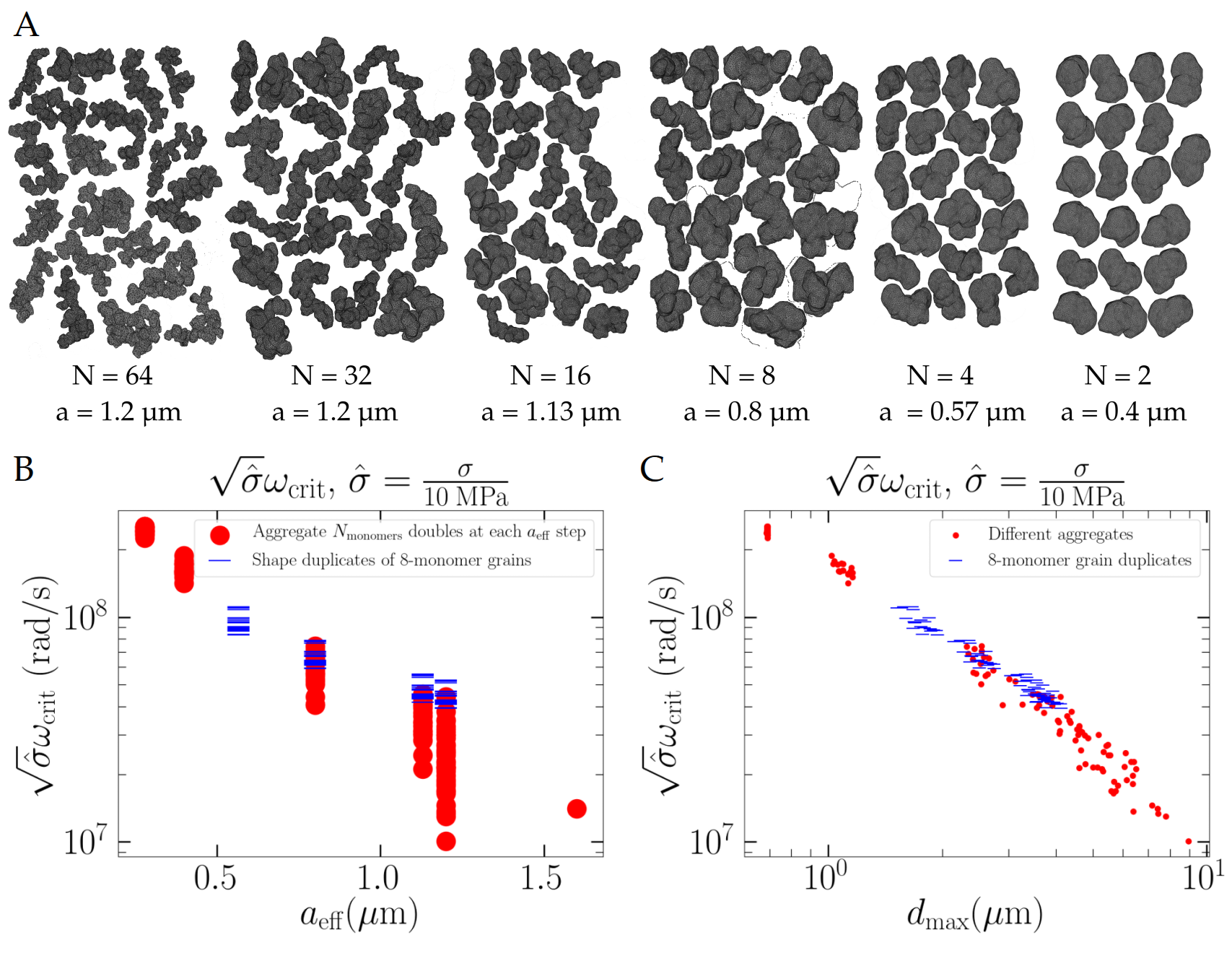}
		\caption{Aggregates used (A) and their critical angular speeds needed for rotational disruption as a function of equivalent sphere radius $a_\mathrm{eff}$ (B) and particle length $d_\mathrm{max}$ (C) with relatively more compact duplicates of $ N=8 $ monomer particles in blue.}
		\label{fig:aggregatesdata}
	\end{figure*}

	The scaling of the ellipsoids is chosen to be such that the monomers overlap, generating a connected aggregate shape. The number of monomers is $ N = 2^{n} $ with $ n = 1, ..., 6 $. The aggregates, illustrated in Fig. \ref{fig:aggregatesdata}A, are scaled so that quadrupling the number of monomers doubles the equivalent sphere radius of the particle, which corresponds to behavior of a fractal with dimension 2, which is known to be close to the fractal dimension of a BCCA particle \citep{Meakin1984}. Accordingly, the monomer radius $ a_{\mathrm{monomer}} $ and the equivalent-volume sphere radius $ a_{\mathrm{eff}} $ are related by \begin{equation}\label{eq:aggrscaling}
	N = \left(\dfrac{a_{\mathrm{eff}}}{a_{\mathrm{monomer}}}\right)^{2}.
	\end{equation} 
	
	Each family of sample aggregates with $ N $ monomers contains 20 distinct shapes. The monomer radius is approximately 0.28 µm for the $ N=1,..,4 $ cases. The largest aggregates have a great discrepancy between their equivalent sphere radii and their lengths, $ d_{\mathrm{max}} $. Lengthy particles and thus large computational bounding boxes impose a challenge to the $ T $-matrix solver, as the effective size parameter and thus the size of the $ T $-matrix is determined from the largest edge length of the particle bounding box. Monomer radii of 0.21 µm and 0.15 µm are chosen for the $ N=32 $ and $ N=64 $ cases, respectively. However, a single particle of $ N=32 $ is scaled to $ a_{\mathrm{eff}} =  1.6$ µm, as a result of a partly successful ensemble construction for the $ a_{\mathrm{monomer}} $ = 0.28 µm case. In addition, for computational efficiency, the $ T$-matrices are computed for only a single wavelength instead of a discretized wavelength continuum.
	
	For the composition of the dust, astronomical silicate with a refractive index $ n = 1.686 + \mathrm{i}0.0312 $ at wavelength $ \lambda = 600  $ nm is used. The $ T $-matrices for all aggregate shapes are computed for the single wavelength $ \lambda = 600 $ nm for less intensive computational requirements, as the largest effective size parameters $ x = 2\pi d_{\mathrm{max}}/\lambda $ are as high as over 10. The energy density of the solar radiation is $ u_{\mathrm{rad}} = 1370/c $ $ \mathrm{W/m^2} $ at 1 au. 
	
	The tensile strength of a spinning dust particle determines whether the centrifugal stresses cause disruption. The centrifugal stress is given by the integral\begin{equation}\label{eq:centrifugal_stress}
	\begin{aligned}
	\mathbf{F}_{c} &= \int\mathrm{d}\mathbf{F}_{c} = -\int\mathrm{d}m \bm{\omega}\times(\bm{\omega}\times\mathbf{r}) \\
	&= -\rho\omega^{2}\int\mathrm{d}V\hat{\mathbf{n}}\times(\hat{\mathbf{n}}\times\mathbf{r}) \equiv -\rho\omega^{2}\sigma_{V},
	\end{aligned}
	\end{equation}
	where $ \rho $ and  $ \omega $ are the particle density and angular speed, respectively, and in the last step the assumption of particle spinning about the axis $ \hat{\mathbf{n}} $ parallel to the major inertial axis is made. The integral is denoted by $ \sigma_{V} $, as it gives the shape-dependent contribution to the centrifugal stress. All particle geometries are discretized as tetrahedral meshes (see Fig. \ref{fig:aggregatesdata}A), allowing straightforward evaluation of the integral $ \sigma_{V} $.
	
	The ultimate tensile strength $ T $ of an aggregate dust particle is estimated by finding the minimum geometric cross section $ A_{\mathrm{min}} $ of the particle along the long axis as $ T = \sigma_{f}A_{\mathrm{min}} $, where $ \sigma_{f} $ is the specific tensile strength of the material. The minimum geometric cross section is determined from the shape discretization inside radius $ d_{\mathrm{max}}/4 $ from the particle center of mass, where $ d_{\mathrm{max}} $ is the particle length along its long axis. The restriction is a simplified means to account for the reduction of centrifugal load when moving outwards from the axis of rotation towards the either end of the particle. Thus, the critical disruption speed will be estimated by \begin{equation}\label{eq:critical_speed}
	\omega_{\mathrm{crit}} = \sqrt{\frac{\sigma_{f}A_{\mathrm{min}}}{\rho|\sigma_{V}|}}.
	\end{equation}
	
	The critical angular speed needed to break the aggregate is illustrated in Fig. \ref{fig:aggregatesdata}B. The critical speed is compared between the aggregates and upscaled duplicates of the aggregates with $ N = 8 $ monomers. The shape duplicates are more compact than the aggregates, which results in higher critical speeds than for the aggregates. In addition, the critical speed is inversely proportional to $ d_{\mathrm{max}} $, in Fig. \ref{fig:aggregatesdata}C. For convenience, the critical speed is normalized with respect to the square root of the specific tensile strength of 10 MPa, $ \hat{\sigma} $. If tensile strength values of either 100 kPa or 1 GPa are considered, the critical speeds would then be either divided or multiplied by 10, respectively.
	
	\subsection{Spin-up by radiative torques}
	Radiative torques can be written in terms of the total electromagnetic field, usually mathematically separated into the incident and scattered fields \citep{Marston1984, Farsund1996}. The RAT $ \mathbf{\Gamma} $ is usually written in terms of the scatterer-dependent torque efficiency $ \mathbf{Q}_{\Gamma} $ as \begin{equation}\label{eq:radiativetorque}
	\mathbf{\Gamma}_{\mathrm{rad}} = \frac{1}{2} u_{\mathrm{rad}} \lambda a_{\mathrm{eff}}^{2} \mathbf{Q_{\Gamma}},
	\end{equation} where $ u_{\mathrm{rad}} $, $ \lambda $, and $ a_{\mathrm{eff}} $ are the radiation energy density, wavelength and the equivalent volume sphere radius of the scatterer. 
	
	\citet{Herranen2018} showed via explicit integration that RATs rapidly cause stable angular speeds in interstellar-like conditions. When interstellar conditions, specifically the radiation intensity, are translated to correspond the circumsolar environment, the timescales of reaching stability are of seconds, corresponding to the million-fold increase in radiation density.
	
	When the stable spin is assumed to occur about the axis of major inertia, which is the most likely axis when the rotational energy is minimized, the radiative torque efficiency $ \mathbf{Q_{\Gamma}} $ can be decomposed into the spin-up, alignment, and precession components $ H $, $ F $, and $ G $ \citep{Draine1997}. 
	
	In the general case of circumsolar dust, where the radiative torque is allowed to change with the heliocentric radius, the angular momentum of a dust particle evolves according to the equation of motion \citep{Hoang2019}\begin{equation}\label{eq:omegaevolution}
	\frac{I_{3}\mathrm{d}\omega}{\mathrm{d}t} = \Gamma_{H}-\frac{I_{3}\omega}{\tau_{\mathrm{damp}}},
	\end{equation} 
	where $ I_{3} $ is the major principal moment of inertia, $ \Gamma_{H} $ is the spin-up component of the radiative torque, and $ \tau_{\mathrm{damp}} $ is the characteristic rotational damping time, and the assumption of stable spin about the major axis of inertia is made. The damping time is the reciprocal sum of characteristic times of different angular velocity constraining phenomena, such as gas drag and loss of angular momentum by IR emission. To consider only the spin-up effect of radiative torques, the accelerating contribution of IR emission is omitted, as are all other possible accelerating torques. If the RAT magnitude is held constant, the solution of Eq. \eqref{eq:omegaevolution} is simply the exponential approach towards a final angular velocity $ \omega_{\mathrm{max}} $, given by\begin{equation}\label{eq:omega_max}
	\omega_{\mathrm{max}} = \frac{\Gamma_{H}\tau_{\mathrm{damp}}}{I_{3}}.
	\end{equation}
	
\begin{figure}
		\centering
		\includegraphics[width=0.9\linewidth]{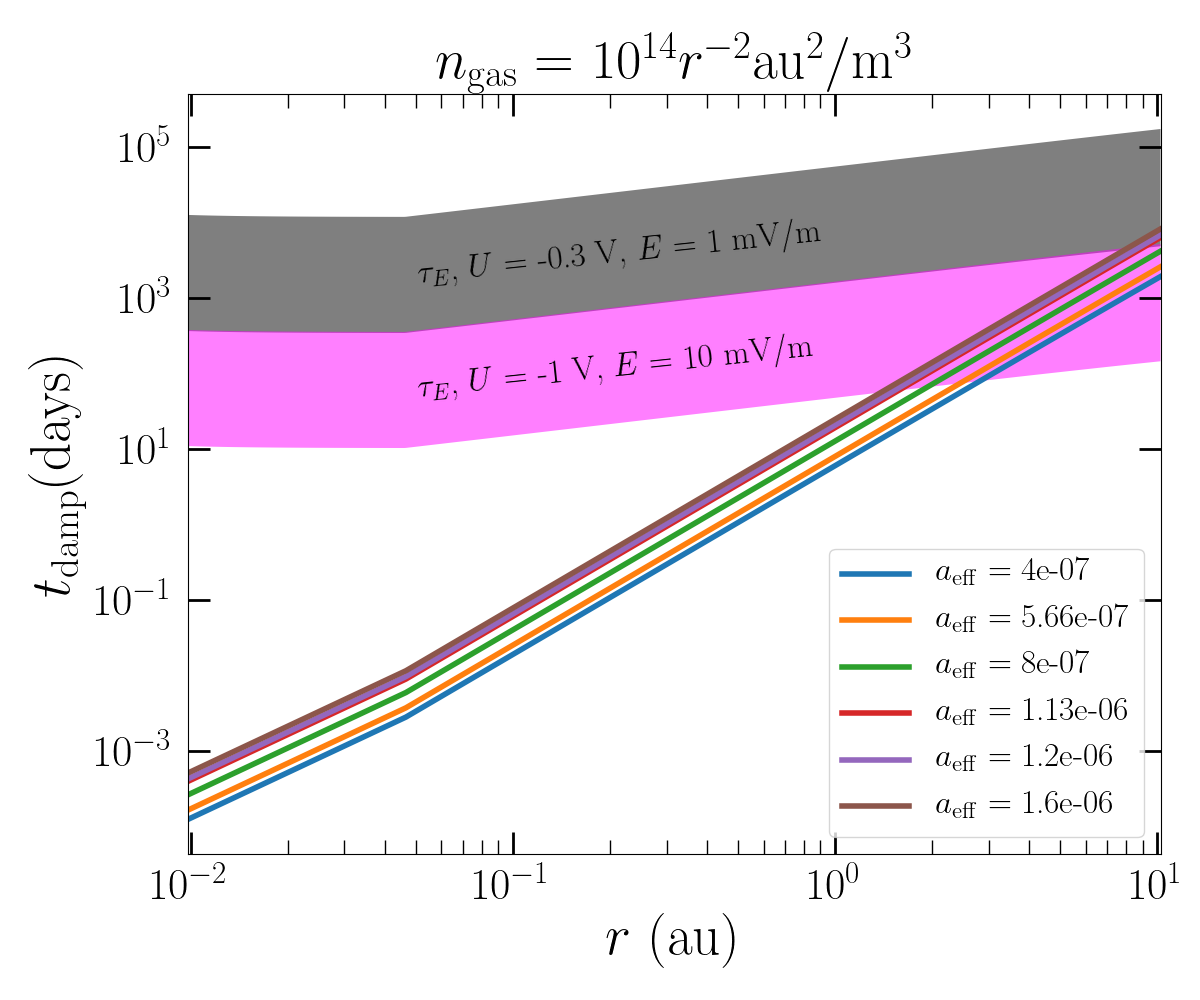}
		\caption{Damping time as a function of heliocentric distance. Electric precession timescale ranges are illustrated by the shaded areas.}
		\label{fig:tdamp}
	\end{figure}

	The damping time $ \tau_{\mathrm{damp}} $ is taken to be the reciprocal sum of gas drag $ \tau_{\mathrm{gas}} $ and IR emission damping $ \tau_{\mathrm{IR}} $, the latter of which can be related to the former as \citep{Draine1996a, Draine1998}:\begin{equation}\label{eq:taus}
	\tau_{\mathrm{damp}}^{-1} = \tau_{\mathrm{gas}}^{-1}+\tau_{\mathrm{IR}}^{-1} = \tau_{\mathrm{gas}}^{-1} + F_{\mathrm{IR}}\tau_{\mathrm{gas}}^{-1},
	\end{equation}
	where 
	\begin{equation}\label{eq:tau_gas}
	\tau_{\mathrm{gas}} = \frac{3}{4\sqrt{\pi}}\frac{I_{3}}{1.2n_{\mathrm{gas}}m_{\mathrm{gas}}v_{\mathrm{th}}a_{\mathrm{eff}}^{4}},
	\end{equation}
	\begin{equation}\label{eq:FIR}
	F_{\mathrm{IR}} \approx 1.2\cdot 10^{-3}U_{\mathrm{rad}}^{2/3}\left(\frac{1\:\mathrm{\mu m}}{a_{\mathrm{eff}}}\right)\left(\frac{10^{14}\:\mathrm{m^{-3}}}{n_{\mathrm{gas}}}\right)\left(\frac{200\:\mathrm{K}}{T_{\mathrm{gas}}}\right)^{1/2}, 
	\end{equation}
	and $ n_{\mathrm{gas}} $ is the number density of the gas, $ m_{\mathrm{gas}} $ the mass of a gas molecule, $ v_{\mathrm{th}} = (2k_{B}T_{\mathrm{gas}}/m_{\mathrm{gas}})^{1/2}$ the thermal speed of the gas, and $ U_{\mathrm{rad}} = u_{\mathrm{rad}}/u_{\mathrm{rad}}(r=\mathrm{1\: au})$ is the normalized radiation intensity. The damping time under these assumptions is plotted for single random representative particles from each size group in Fig. \ref{fig:tdamp}. The IR damping, which under the assumptions here would result in a fan-like spread of the damping time, is rather insignificant compared to gas drag, as is also evident from Eq. \eqref{eq:FIR}. The bends on the curves are due to the gas temperature being held at the maximum temperature of 3000 K, approximately the sublimation temperature of silicate.
	
	\subsection{Cometary environment}
	Environment plays a key role in whether or not rotational disruption is possible. Specifically, the interplay of radiation intensity, variable gas density, IR damping, shielding of dust inside an optically thick coma, solar wind conditions, magnetic fields, etc, create an extraordinarily complicated whole, where the fate of one single dust particle might be wildly different from that of another.  It is difficult to analyse RAT properties in all possible environments. For maximally general conclusions to be drawn, the most important competing effects need to be parametrized as simply as reasonably possible. Further, a conservative estimation of the RAT significance against the damping effects present allows comparison with effects omitted in the work, such as magnetic fields outside the coma, particle collisions, extreme gas flows or sublimative torques. 
	
	First, to simplify the picture, dust response in flowing gas with a number density $ n_{\mathrm{gas}} $ with IR damping is taken into account when considering the disruption process. Alignment depends on the interplay of the timescales of relaxation due to different physical phenomena, such as interactions with external magnetic or electric fields. Precession of particles with an electric dipole moment can occur in an electric field produced by anisotropic plasma or by relative motion in a magnetic field, however, Larmor precession about the magnetic field is insignificant in the coma \citep{Goetz2016, Hajra2018}. The electric precession time $ \tau_{E} $ is given by
	\begin{equation}\label{eq:E_precession}
	\begin{aligned}
	\tau_{E} = &0.08\left(\dfrac{\rho}{3000 \mathrm{\: kg/m^3}}\right)\left(\dfrac{U}{0.3 \:\mathrm{V}}\right)^{-1}\left(\dfrac{E}{1\: \mathrm{mV/m}}\right)^{-1} \\
	&\times\left(\dfrac{\epsilon}{0.01}\right)\left(\dfrac{\omega_{\mathrm{max}}}{\omega_{\mathrm{th}}}\right)\left(\dfrac{T_{\mathrm{gas}}}{100\:\mathrm{K}}\right)\left(\dfrac{a_{\mathrm{eff}}}{1\:\mathrm{\mu m}}\right)^{1/2} \:\mathrm{days}.
	\end{aligned}
	\end{equation}
	Above, $ \rho $, $ \epsilon $, and $ U $ are the particle density, charge distribution asymmetry, and electrostatic potential, respectively. $ E $ is the external electric field magnitude, and $ \omega_{\mathrm{th}} = (2k_{B}T_{\mathrm{gas}}/I_{3})^{1/2} $ is the thermal angular velocity. Using $ \omega = 0.1\omega_{\mathrm{max}} $, the electric precession time is illustrated by the shaded area in Fig. \ref{fig:tdamp}. When the angular speeds are suprathermal, especially when close to the maximum speed due to the RATs, the electric precession is insignificant, or $ \tau_{E}>\tau_{\mathrm{damp}} $, during the active cometary phase.
	
	The number density of the gas around a dust particle depends linearly on the gas release flux and distance of the dust from the cometary nucleus according to the inverse-square law. For simplicity, only the release flux, which also has an inverse square dependence on the heliocentric distance $ r_{H} $ \citep{Sanzovo2001}, is taken to affect the number density. Also, the comet is assumed to be homogeneously active. Thus, the gas number density is given by\begin{equation}\label{eq:n_gas}
	n_{\mathrm{gas}} = n_{0}r_{H}^{-2},
	\end{equation}
	where $ n_{0} = 10^{14} \: \mathrm{m^{-3}}$ at 1 au at some unspecified distance from the nucleus. The number density is expected to fall similarly according to a power law when the distance to the nucleus increases. While highly relevant when the whole coma is modeled, the detailed modeling of the gas distribution falls out of the scope of the study.
	
	Gas drag depends on the thermal velocity of the gas. The gas is taken to be H\textsubscript{2}O, the most common volatile available inside $ r_{H} \approx $ 3 au. The free sublimation temperature of water, according to its phase diagram, is approximately 200 K. The temperature can be regarded as the lower limit for the gas temperature. However, it is likely that the gas is released with higher temperatures due to contact with the cometary surface, assumed to be in black body equilibrium. The equilibrium temperature provides an upper limit to the gas temperature, which in turn determines the lower limit for both $ \tau_{\mathrm{gas}} $ and the relative RAT significance. 
	
	\section{Spin-up and disruption of aggregate particles}
\begin{figure}
		\centering
		\includegraphics[width=0.9\linewidth]{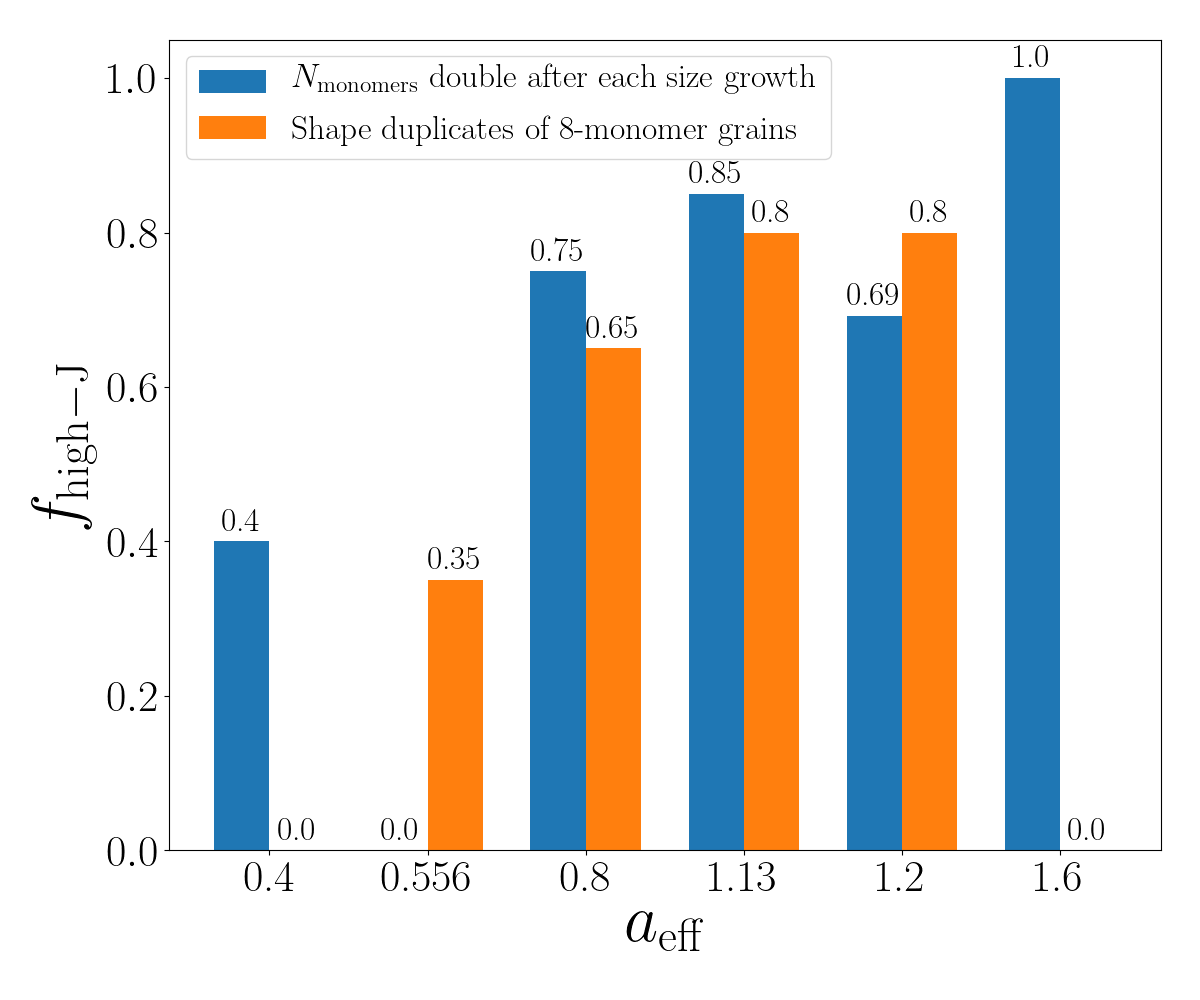}
		\caption{Fraction $f_{\mathrm{high-}J}$ of aggregate particles (blue) and the more compact shape duplicates of $ N=8 $ monomer particles (orange).}
		\label{fig:fhighj}
	\end{figure}

	Radiative torques cause alignment addition to the spin-up. Alignment can be quantified using a single parameter, the fraction $ f_{\mathrm{high-}J} $ of particles for which there exists an attractor point with a high value of angular momentum $ J $ \citep{Hoang2008}. The fraction $ f_{\mathrm{high-}J} $ of the dust particles, whose dynamics is dominated by RATs, is assumed to be aligned in the characteristic damping time $ \tau_{\mathrm{damp}} $. In Fig. \ref*{fig:fhighj}, it is seen that the expected aligned fraction increases when particle sizes grow, consistent with the canonical properties of RATs predicting that smaller-than-wavelength particles are ever-decreasingly aligned when particles become even smaller \citep{Lazarian2007b}. The existence of high values of $ f_{\mathrm{high-}J} $ provide necessary, but not sufficient conditions for polarization by aligned dust to occur, as strongly precessing particles can appear de-aligned although the angular momentum is fixed.
	
\begin{figure*}
		\centering
		\includegraphics[width=0.9\linewidth]{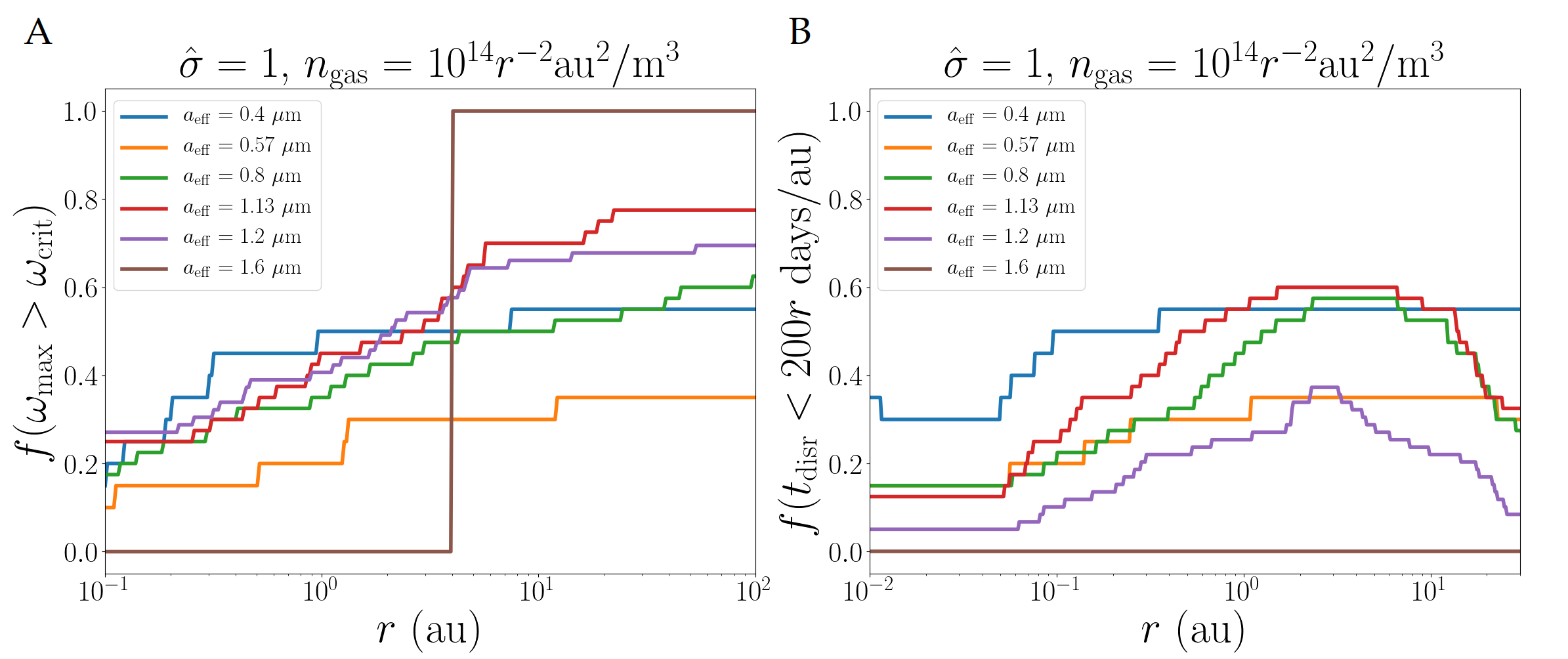}
		\caption{Fraction of aggregate particles that can reach disruptive angular speeds (A) and that reach the disruptive speeds in circular orbits in timescales close to those that a comet in Keplerian orbit spends at a certain distance (B).}
		\label{fig:disruption}
	\end{figure*}

	Not all particles, even those with a high-$ J $ attractor, are guaranteed to reach critical angular speeds. Next, the fraction of particles in circular orbits, for which the disruption criteria are satisfied, are considered. For simplicity, the high-$ J $ attractor of lowest value of $ J $ is considered for any particle with multiple attractors, yielding a lower limit on the amount of disrupted particles. In Fig. \ref{fig:disruption}A, the fraction of particles that have a critical maximum angular speed, according to Eqs. \eqref{eq:critical_speed} and \eqref{eq:omega_max}, is shown to be between 20 and 50 \% at 1 au. 
	
	It is unclear whether or not dust in Keplerian orbits can reach disruptive angular speeds during cometary activity. In Fig. \ref{fig:disruption}B, a time scaling of 200 days times the heliocentric distance is used in circular orbits. The scaling emulates elliptic orbits, where comets are in their closest vicinity to the Sun from days up to weeks. It is shown that the largest particles are not all accelerated quickly enough, even if their maximum angular speed were disruptive. Ever-larger particles are unlikely to reach disruptive speeds either, as a constant RAT magnitude is expected for very large particles, while their inertia still increases.
	
\begin{figure*}
		\centering
		\includegraphics[width=0.9\linewidth]{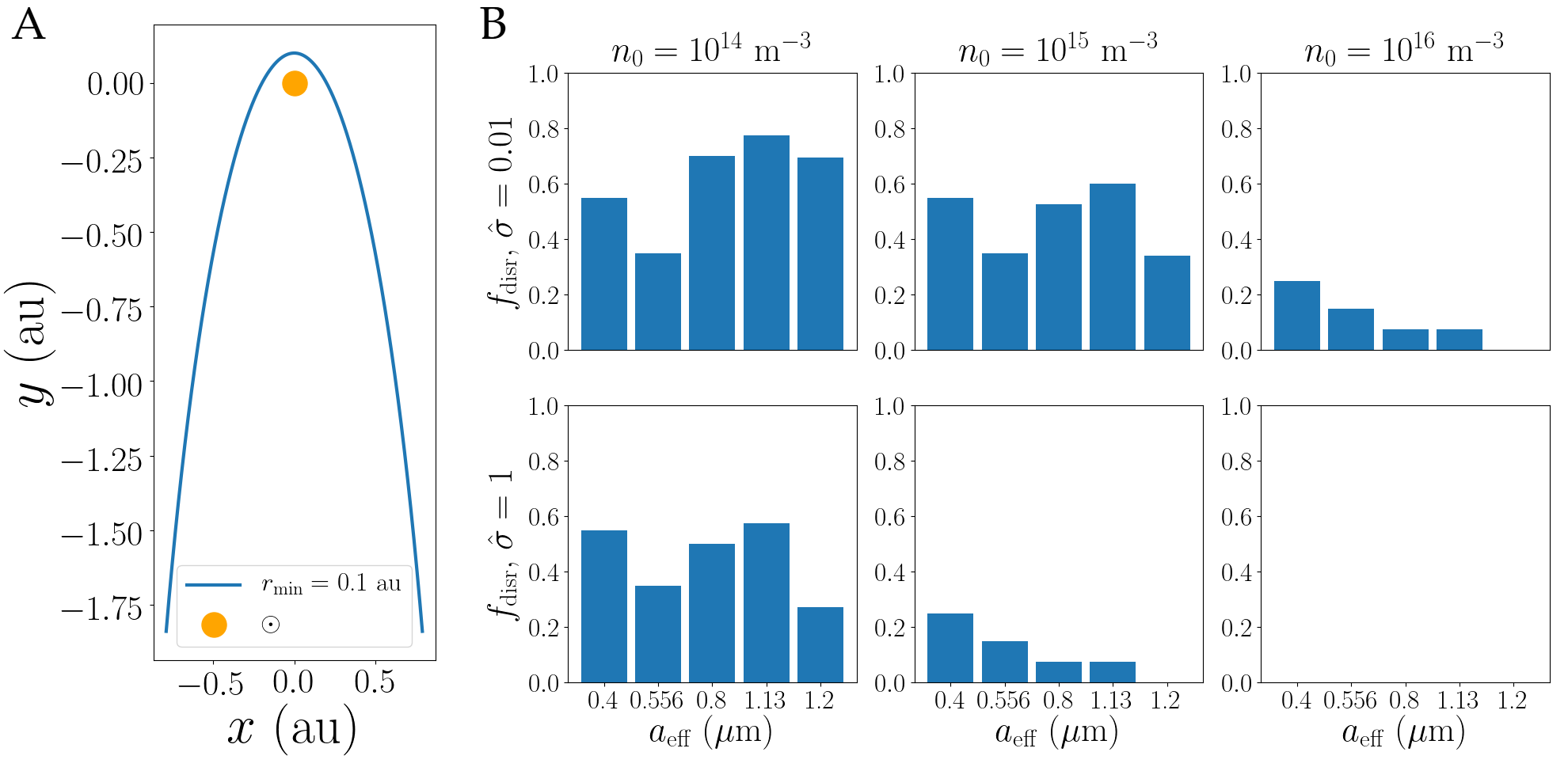}
		\caption{(A) Orbit of a near-Sun comet with a semi-major axis of 5 au inside the heliocentric radius of 2 au, assumed to be the region of cometary activity. (B) Disrupted fractions of particles assuming different tensile strengths and gas densities during the active phase of the comet.}
		\label{fig:disruption_statistics}
	\end{figure*}

	The total disrupted fraction $ f_{\mathrm{disr}} $ of the ensemble is now estimated for a near-Sun comet with a perihelion $ r_{p} =$ 0.1 au and a semi-major axis $ a= $ 5 au with varying $ \hat{\sigma} $ and $ n_{0} $. The comet, the orbit of which is visualized in Fig. \ref{fig:disruption_statistics}A, is active inside heliocentric radius $ r= $ 2 au. A gas flow described by Eqs. \eqref{eq:taus} and \eqref{eq:n_gas} embeds the dust and each particle is initially at rest. 
	
	The disrupted fraction, illustrated in Fig. \ref{fig:disruption_statistics}B, is consistent with the results presented in Fig. \ref{fig:disruption}. Particles of size $ a_{\mathrm{eff}} = $ 0.56 µm are the most weakly disrupted when the gas density and/or tensile strength are relatively low. The RAT disruption can thus produce an excess of 0.56 µm particles, unless larger aggregates have the same relatively high value of $ \hat{\sigma} $ as smaller ones or the damping effect of higher gas density dominates. The former, however, is unlikely under the assumption that smaller particles are more compact and thus have higher tensile strengths. Rotational disruption does not occur if the gas density and tensile strength are both high enough.
	
	\subsection{Spectral bluing by disruption processes}\label{sec:bluing}
	\citet{Brown2014} describes the conditions for spectral bluing (as opposed to reddening, or a red spectral slope) for small Mie scatterers. In their work, it is found that the optimal range of the size parameter $ x = ka_{\mathrm{eff}} = 2\pi a_{\mathrm{eff}}/\lambda$ for Mie spheres where spectral bluing universally occurs for a range of different optical properties is $ x \in 0.5-1.2 $. In the near-infrared range of 1.25 to 2.25 microns, corresponding to an excess of dust particles in the size range of about 0.1 to 0.4 microns.
	
	\citet{Petrova2000} note that when nonspherical aggregates are considered, straightforward application of extinction results derived from spheres is likely to underestimate the particle size. The size range responsible for bluing may thus well be larger than the 0.1--0.4 µm range, which would be more consistent with results in Fig. \ref{fig:disruption_statistics}B in certain conditions.
	
	When particles are disrupted, smaller particles emerge. Here, assuming BCCA particles, the disruption is assumed to simply produce two subparticles half the size of the parent particle. Then, the complement of the disrupted fraction gives the population that survives the RAT disruption process, and the disrupted fraction gives half of the amount of particles half the original size that are produced. Next, it is assumed that the disrupted fraction will be zero for very small particles due to weaker RATs and zero for very large particles due to their ever-larger inertia and constant RATs. A simple spline is fitted to the disrupted fraction using the zero boundary condition assumptions above. 
	
	In the following, a simplified picture of the disruption process is adopted. Namely, disruption occurs simultaneously and strictly once in each size group. The simple picture contrasts with a more realistic model, where different disruption timescales between size groups are taken into account, affecting both the production of particles with an initial angular velocity $ \omega $ and their disruption time scales over the whole size distribution, i.e. conditions for so-called fragmentation cascade. 
	
\begin{figure}
		\centering
		\includegraphics[width=\linewidth]{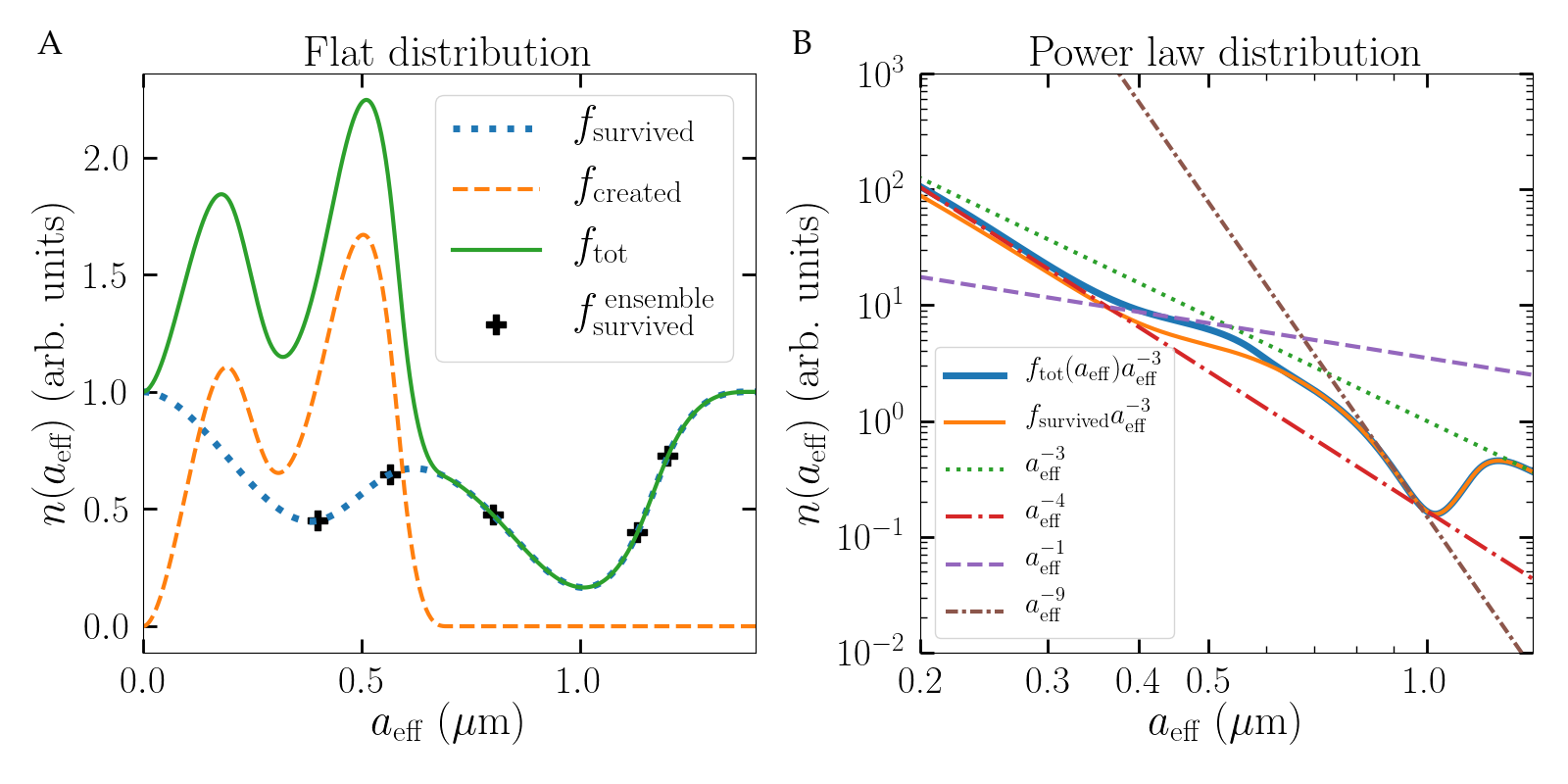}
		\caption{A: Effect of disruption on a flat size distribution, based on a spline fit over the actual numerical RAT disruption results. B: The population change, taking account either the depletion by disruption and creation from parent particle, or just the depletion, when a power law size distribution with a power law index $ -3 $ is assumed, with power law fits over different parts of the distribution.}
		\label{fig:disruptedfraction}
	\end{figure}

	In Fig. \ref{fig:disruptedfraction}A, the change in a flat size distribution under the simplified picture is illustrated. The 0.5-µm particles are expected to have the greatest excess, with the 0.2-µm excess being a close second. The 1-µm population is the most depleted one. 
	
	Next, as an example, a differential power law size distribution with an exponent of $ -3 $ is assumed to describe an initial dust population. Similar analysis can be made for any initial size distribution, starting from the fraction of survived particles from Fig. \ref{fig:disruptedfraction}A. The disruption process significantly affects the size distribution in the 0.4--1.0 µm range, where the power law index locally changes drastically from the initial value. The depletion of particles is, as expected under an assumption of a power law distribution, the dominant factor of altering the distribution. Although the production of new particles produces a significant absolute excess, the distribution shape is weakly affected. In Fig. \ref{fig:disruptedfraction}B, it is seen that the disruption depletion changes the overall trend of the power law index from $ -3 $ to $ -4 $, with the steepest index of $ -9 $ near the most depleted 1 µm size group.
	
	\citet{Steckloff2016} consider fragmentation cascade in the context of fragmentation of macroscopic chunks due to sublimation-driven YORP effect. Similar cascade-like fragmentation of the child particles with a considerable initial angular speed compared to the critical angular speed is a physically reasonable model, where the dependence of the disruption timescale on the particle size and tensile strength is of key interest.
	
	Until now, the tensile strength of the ensemble has been held constant over the whole size range. However, single monomers are likely compact grains that are strongly held intact. Larger aggregates are likely weaker. In macroscopic objects, tensile strength decreases as a function of size due to microscopic cracks as $ \sigma \sim a_{\mathrm{eff}}^{-1/2} $ \citep{Griffith1921}. In microscopic aggregates, a contact point can be either weakly bound by e.g. Van der Waals adhesion or more strongly, if the contact point actually is a part of a single crystalline structure. When monomer number increases, less likely it is for a whole aggregate to be a single crystalline particle.  
	
\begin{figure}
		\centering
		\includegraphics[width=\linewidth]{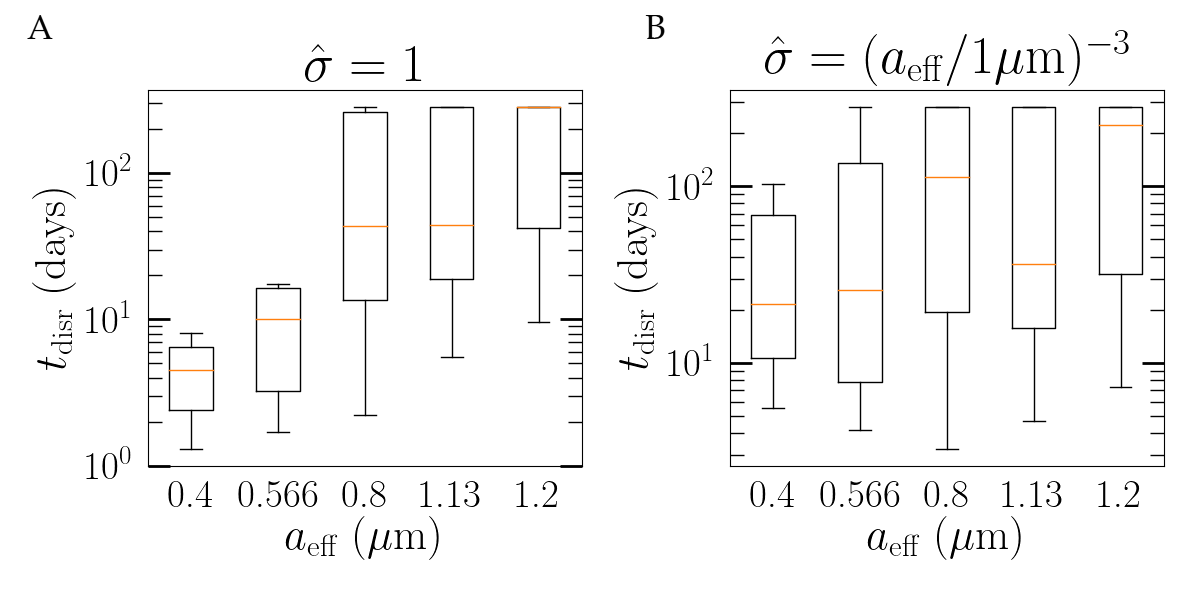}
		\caption{Box-whisker plots of the disruption timescales of a constant tensile strength ensemble (A) and an ensemble with a size-dependent tensile strength (B).}
		\label{fig:timescalechange}
	\end{figure}

	As an example model, adopting a simple size dependence on the tensile strength normalization of $ \hat{\sigma} = (a_{\mathrm{eff}}/1$ µm$ )^{-3} $ results in a smooth transition from $ \hat{\sigma} = $ 15 for the 0.4 µm case to $ \hat{\sigma} = $ 0.6 for the 1.2 µm case. For comparison, the size-dependent tensile strength of micron-to-millimeter-scale aggregates follows, according to numerical simulations \citep{Tatsuuma2019}, a $ \hat{\sigma} = (a_{\mathrm{eff}}/1$ µm$ )^{-2} $ law. The example model overestimates tensile strengths given by both the microcrack theory and numerical simulations of aggregates, as the smallest particles are composed of two monomers. However, if the smallest particles were crystalline, the model underestimates their strength. A reasonable estimate for the strength of a crystalline material would be given by $ \hat{\sigma} = 100 $, as tabulated in \citep{Hoang2019}. In Fig. \ref{fig:timescalechange}, the constant tensile strength case (A) is compared against the size-dependent case (B) for dust around a near-Earth asteroid with a perihelion distance $ r_{p} = 0.99$ au and an aphelion distance $ r_{a} = 5.1 $ au inside the heliocentric distance of 2 au. 
	
	The time of disruption depends on the critical angular speed and the spin-up time, the former of which is a function of the tensile strength. When the tensile strength is constant, the smaller particles have far lower disruption timescales than the larger ones. If the tensile strength is dependent on the particle size, as to some limit it is plausible to assume to be true, disruption timescales become more uniform. Thus, the results of Fig. \ref{fig:disruptedfraction}B are more representative, if the smaller particles are more difficult to be disrupted. The disruption timescales roughly follow similar scaling than the critical angular velocity, which can be scaled according to the specific tensile strength, as in Eq. \eqref{eq:critical_speed}. 
	
	Finally, if size-dependent tensile strength is assumed, the disruption timescales can become more uniform. Then, Figs. \ref{fig:disruptedfraction} A and B provide insight on how fragmentation-cascade-corrected dust size distribution can vary. The depletion of efficient scatterer of red and NIR wavelengths implies spectral bluing in scattered light compared to a pristine dust population. In addition, the largest-sized excess produced by disruption is likely to overlap with the size range where bluing by non-spherical particles occurs, according to Fig. \ref{fig:disruptedfraction}A. However, depending on the disruption time scale differences of differently-sized particles and whether or not a fragmentation cascade occurs in the size range of bluing particles, the bluing particles may be removed from the population even more efficiently than they are created. Thus, the RAT disruption effect on the spectral colour is likely closely linked to the characteristic size and tensile strength of the smallest solid crystalline constituent of the aggregate.
	
	\subsection{Alignment and polarization}\label{sec:alignment}
	Disruption is assumed to occur if, for a single particle, there exists a mode of high angular momentum in a $ (J,\Theta) $ space \citep[see][]{Hoang2008}, where the direction of angular momentum $ \Theta $ is measured with respect to the incident radiation direction. Internal relaxation effects \citep{Hoang2009} eventually force the particle to a state where the major axis of inertia is parallel to the angular momentum. 
	
\begin{figure}
		\centering
		\includegraphics[width=0.9\linewidth]{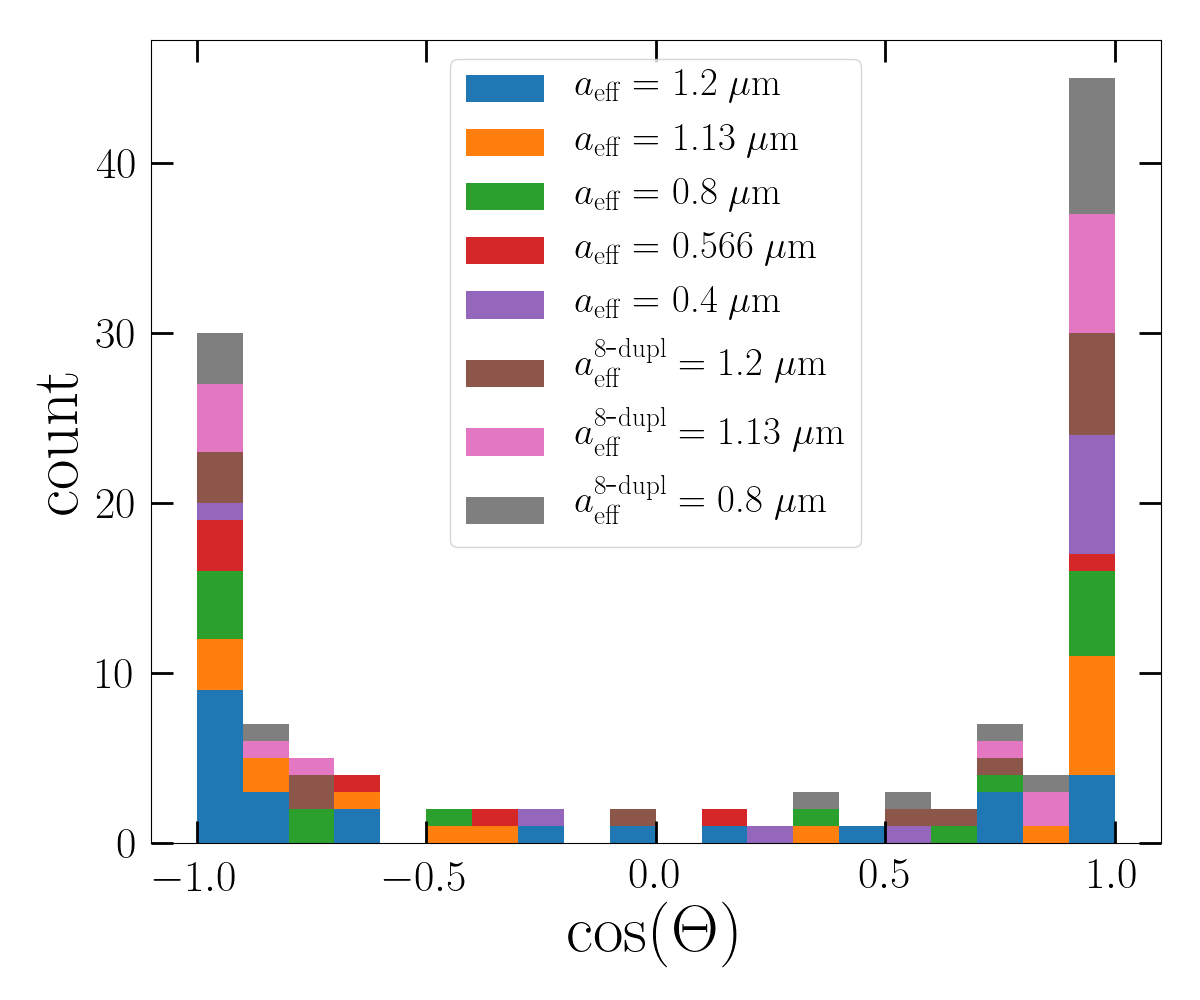}
		\caption{Distribution of the high-$ J $ attractor directions most parallel to the incident light direction of the particle ensemble.}
		\label{fig:alignmentangle}
	\end{figure}

	The angular momentum is free to precess around the incident light direction. Unless $ \cos\Theta=\pm 1 $, the alignment is not perfect, and the effect on polarization is weaker. Alignment with respect to the incident light direction is called $ k $-RAT alignment and it occurs in timescales given by Eq. \eqref{eq:taus}. In Fig. \ref{fig:alignmentangle}, the distribution of directions of high-$ J $ attractors of all differently sized particles are collected. If there exists multiple attractors, only the one most parallel to $ k $ is considered, marking the preferred direction when randomizations are present. 
	
	Alignment can also occur with respect to magnetic fields, in which case it is referred to as $ B $-RAT alignment. The effect of magnetic fields is disregarded here, while electric fields are considered in Fig. \ref{fig:tdamp}. In cometary comae, electric fields can align particles for which high-$ J $ attractors do not exist, i.e., the complement of particles composing the fraction $ f_{\mathrm{high-}J} $ in Fig. \ref{fig:fhighj}. 
	
	In general, when competing RAT mechanisms exist simultaneously, the one with the shortest characteristic timescale will determine the axis of precession. When the conditions impose different RAT effects of comparable scale on the particles simultaneously, particles will precess more erratically and the degree of alignment is likely lower than when either of the effects dominate. 
	
	When dust is released to the coma and the dust tail, $ k $-RAT alignment can produce a large amount of aligned particles quickly. The damping timescales are, inside 1 au, in the order of hours or days. Unless electric fields are extraordinarily strong, the gas damping makes $ k $-RAT alignment preferable, as is seen in Fig. \ref{fig:tdamp}. Additionally, parallel $ J $ and $ k $ is almost guaranteed according to Fig. \ref{fig:alignmentangle}. According to \citet{Hoang2014}, the electric field alignment is predominant after a certain distance from the cometary nucleus. The change in alignment processes is due to the fact that the gas damping time will increase as the gas density decreases.
	
	In essence, the RAT alignment and disruption processes are included in the life of cometary dust as follows. First, inside the gas-dense part of the coma, fast alignment due to $ k $-RATs is possible, and for the ensemble considered here, also very likely. As gas density decreases and $ k $-RAT alignment becomes subdominant, a single dust particle can be accelerated to even higher angular speeds, possibly leading to the disruption of the particle far enough of the nucleus, which would contribute to the fading of cometary tails.
	
	\section{Discussion}
	It was found that RATs can cause rotational disruption of cometary dust within the timescales of cometary activity. The study was done assuming parameter values that are common in estimations of different cometary environments based on observational and measured evidence. As the behavior of comets is highly variable, the results presented here have different direct and indirect implications on the disruption process in the case of any particular comet. For example, the gas-to-dust production and density ratios as varying both in time and distance from the nucleus not only directly affect the disruption process, but are from the fundamental level complicated to account for the coma as a whole \citep{Marshall2019}. Still, the results presented here provide computational evidence for the relevance of radiative torques in observations of cometary dust.
	
	Spin-up and disruption by radiative torques were shown to occur for a wide range of gas densities and tensile strengths. Thus, RATs are likely an affecting factor in all kinds of comets, either as a cause of destruction or polarization of dust. Spin-up causes a gradually increasing degree of polarization. The increase of polarization degree occurs according to the damping timescales presented in Fig. \ref{fig:tdamp}.
	
	The numerical results were consistent with multiple observations, as RATs can in some environments cause spectral bluing or quick increases of polarization degree of near-infrared (NIR) observations by the disruptive or spin-up-combined-with-alignment processes. Even in the case where single observations can be explained by another dominating factor, such as dust destruction by sublimation in the case of sungrazing comets or the effects of solar wind or plasma on the dynamics of dust, the parameter ranges where RAT effects may occur are too large for them to be dismissed solely by a single competing explanation.
	
	All results presented earlier can be summarized as follows:
	\begin{enumerate}
		\item The BCCA particle ensemble of 120 particles are confirmed to have critical maximum angular speeds inversely proportional to their size with compactness as an affecting factor.
		\item The gas damping time is short enough for the $ k $-RAT alignment to occur within the timescale of cometary activity.
		\item The fraction of possibly strongly aligned particles with large angular momenta is large and falls with the particle size, which agrees with established properties of RATs.
		\item Disruption due to RATs occurs when gas densities are not too high and radiation density is high enough, and disruption most efficiently occurs within the distance of high cometary activity.
		\item Fraction $ f_{\mathrm{high-}J} $ correlates with the disrupted fraction $ f_{\mathrm{disr}} $, with 20 to 50 \% of the particles being disruptable depending on their size.
		\item The size distribution and corresponding power laws are modified by the disruption process in the size range of efficient scatterers in the visible and near infrared wavelengths. Size-dependent tensile strengths allow straightforward implication of the results from analysis of $ f_{\mathrm{high-}J} $.
		\item The BCCA ensemble is efficiently and larger particles almost perfectly aligned in terms of $ k $-RAT alignment when environmental conditions allow dominating RATs.
	\end{enumerate}
	
	With respect to the RAT theory, especially when considering the post-perihelion phase of near-Sun comets, where comets regain a dust coma and a dust tail, the future of observations is exciting. In the case where good observational data of reactivating comets can be obtained, new means of not only observational tests of the RAT theory, but tests by in-situ measurements, may be possible. Evidence supporting or contracting RAT disruption in different environments will likely also provide information on how fundamental electromagnetic interactions with radiation contribute to the environment-dependent evolution of dust particles.


\end{document}